\documentclass[12pt,draftclsnofoot, onecolumn]{IEEEtran}
\IEEEoverridecommandlockouts
\usepackage{cite}
\usepackage{amsmath,amssymb,amsfonts}
\usepackage{algorithmic}
\usepackage{graphicx}
\usepackage{textcomp}
\usepackage{xcolor}
\usepackage{algorithm}
\usepackage{algorithmic}
\usepackage{float}
\usepackage{subfigure}
\usepackage{enumerate}
\usepackage{stfloats}
\usepackage{bm}

\newtheorem{theorem}{\textbf{Theorem}}

\newtheorem{lemma}{\textbf{Lemma}}

\newtheorem{problem}{Problem}

\def\BibTeX{{\rm B\kern-.05em{\sc i\kern-.025em b}\kern-.08em
		T\kern-.1667em\lower.7ex\hbox{E}\kern-.125emX}}
\begin{document}
	\title{Analysis and Optimization of Outage Probability in Multi-Intelligent Reflecting Surface-Assisted Systems}

\author{Zijian~Zhang,\thanks{The authors are with Shanghai Jiao Tong University, China.
		
	}~Ying~Cui,~Feng~Yang,~and~Lianghui~Ding
}
	\maketitle
	\begin{abstract}
		Intelligent reflecting surface (IRS) is envisioned to be a promising solution for designing spectral and energy efficient wireless systems. In this letter, we study a multi-IRS-assisted system under Rician fading where the phase shifts adapt to only the line of sight (LoS) components. First, we analyze and optimize the outage probability of the multi-IRS-assisted system in the slow fading scenario for the non-LoS (NLoS) components. We also show that the optimal outage probability decreases with the numbers of IRSs and elements of each IRS when the LoS components are stronger than the NLoS ones. Then, we characterize the asymptotically optimal outage probability in the high signal-to-noise ratio (SNR) regime, and show that it decreases with the powers of the LoS components. To the best of our knowledge, this is the first work that studies the outage probability in multi-IRS-assisted systems.
		
	\end{abstract}

	\section{Introduction}
	With the rapid growth of global wireless data traffic, energy consumption in wireless networks is increasing at a very fast rate. Therefore, spectral and energy efficient design is attracting more and more attention. Recently, intelligent reflecting surface (IRS), consisting of nearly passive, low-cost, reflecting elements with reconfigurable parameters, is envisioned to be a promising solution toward this direction. Specifically, the phase shifts of all passive elements at an IRS can be adjusted such that the reflected signals via the IRS can add coherently with the signals over the other paths at the desired receiver to effectively boost the received signal power.
	
 	A significant amount of research effort has been devoted to performance optimization of IRS-assisted wireless communications. For example, in \cite{1, 2,4,3}, the authors focus on the transmission from a source node to one or multiple destination nodes with the help of a single multi-element IRS. In particular, in \cite{1, 2, 4}, it is assumed that the beamforming vector at the source and the phase shifts of the IRS are adaptive to the instantaneous channel state information (CSI). The maximizations of energy efficiency \cite{1}, received signal power \cite{2} and weighted sum-rate \cite{4} are investigated, and alternating iterative methods are adopted to obtain near optimal solutions. In contrast, in \cite{3}, it is assumed that the direct channel between the source and the destination follows Rayleigh fading, while the reflected channel via the IRS is under Rician fading; and the phase shifts of the IRS adapt to only the line of sight (LoS) components of the reflected channel, while and the MRT beamformer at the source can adapt to the instantaneous CSI. The authors obtain the closed-form optimal phase shifts that maximize the ergodic capacity.
 	
 	Note that the instantaneous CSI-adaptive designs in \cite{1,2,4} have higher implementation costs; the LoS-adaptive design in \cite{3} successfully reduces implementation complexity for fast varying channels. It is also highly desirable to obtain a cost-efficient design for slowly varying channels. In addition, notice that \cite{1,2,3,4} consider the scenario with only one IRS. It is still not known how to jointly adjust the phase shifts of multiple IRSs and how the optimal performance of a multi-IRS-assisted system increases with the number of IRSs. 
 	
 	In this letter, we shall shed some light on the aforementioned challenges. We consider a multi-IRS-assisted system where one singe-antenna source node serves one single-antenna destination node with the help of multiple multi-element IRSs. We consider a Rician fading channel model and allow the phase shifts to adapt to only the LoS components, as in \cite{3}. In contrast with \cite{3}, we focus on the slow fading scenario for the non-LoS (NLoS) components. First, we obtain the expression of the outage probability of the multi-IRS-assisted system. Then, we optimize the phase shifts to minimize the outage probability. We also show that the optimal outage probability decreases with the number of IRSs and the number of elements of each IRS when the LoS components are stronger than the NLoS components. Finally, we characterize the asymptotically optimal outage probability in the high signal-to-noise ratio (SNR) regime, and show that it decreases with the powers of the LoS components. The analytic and optimization results offer important design insights for practical multi-IRS-assisted systems.


	\section{system model}

	As shown in Fig. 1, a single-antenna source (s) node serves a single-antenna destination (d) node with the help of $K$ multi-element IRSs. In particular, the $k$-th IRS has $N_k$ elements. Unlike \cite{3}, there are no requirements on the arrangement of the elements for each IRS. The source is on a high-rise building, while the destination is on the ground. Suppose that the locations of the source and destination do not change during a certain period. The IRSs are installed on the walls of some high-rise buildings around the destination. The phase shifts of all elements of each IRS can be adjusted to assist the transmission from the source to the destination. 

\begin{figure}[h]
	\centering
	\includegraphics[width=0.5\textwidth]{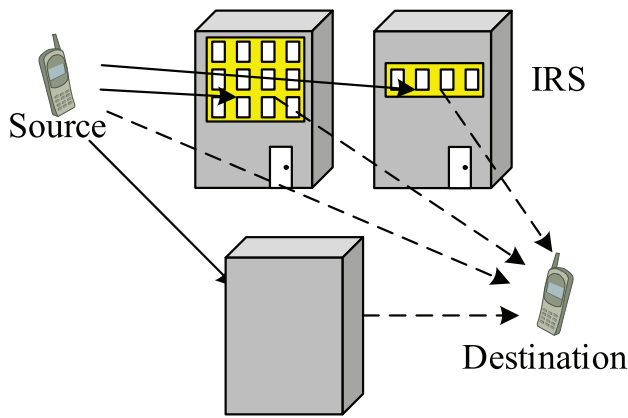}
	\caption{System model. $K = 3$.}
\end{figure}

	Denote $\mathcal{K}\triangleq\{1,\cdots, K\}$ and $\mathcal{N}_k\triangleq\{1,\cdots, N_k\}$, $k\in\mathcal{K}$. For all $k\in\mathcal{K}, n\in\mathcal{N}_k$, the channel between the source and the $n$-th element of the $k$-th IRS, denoted by $h_{k,n}^{(s,r)}\in \mathbb{C}$, is assumed to be a line of sight (LoS) path.\footnote{This assumption is valid, as there is no rich scattering between the source and the IRSs that are all far above the ground and the LoS path is dominant.} In particular, $h_{k,n}^{(s,r)} = \sqrt{\alpha^{(s,r)}_k} \bar{h}_{k,n}^{(s,r)}$, where $\alpha^{(s,r)}_k>0$ represents the distance-dependent path loss and $\bar{h}_{k,n}^{(s,r)}\in\mathbb{C}$ with $|\bar{h}_{k,n}^{(s,r)}| = 1$ represents the normalized LoS path [1], [2]. In addition, assume that the direct channel between the source and the destination, denoted by $h^{(s,d)}\in\mathbb{C}$, and the channel between the $n$-th element of the $k$-th IRS, denoted by $h^{(r,d)}_{k,n}\in\mathbb{C}$, have both LoS and NLoS components.\footnote[2]{This assumption is valid, as scattering is often rich near the ground.} Specifically, $h^{(s,d)}$ and $h^{(r,d)}_{k,n}$ are assumed to follow Rician fading, i.e.,  
	\begin{equation}
	\nonumber
	h^{(s,d)}=\sqrt{\alpha^{(s,d)}}\left(\sqrt{\frac{\kappa^{(s,d)}}{\kappa^{(s,d)}+1}}\bar{h}^{(s,d)}+\sqrt{\frac{1}{\kappa^{(s,d)}+1}}\tilde{h}^{(s,d)} \right),
	\end{equation} 
	\begin{equation}
	\nonumber
	h^{(r,d)}_{k,n}=\sqrt{\alpha_{k}^{(r,d)}}\left(\sqrt{\frac{\kappa_{k}^{(r,d)}}{\kappa_{k}^{(r,d)}+1}}\bar{h}_{k,n}^{(r,d)}+\sqrt{\frac{1}{\kappa_{k}^{(r,d)}+1}}\tilde{h}_{k,n}^{(r,d)}\right), 
	\end{equation}
	where $\alpha^{(s,d)},\alpha_{k}^{(r,d)}>0$ represent the distance-dependent path losses, $\bar{h}^{(s,d)}, \bar{h}_{k,n}^{(r,d)} \in\mathbb{C}$ with $|\bar{h}^{(s,d)}|,|\bar{h}_{k,n}^{(r,d)}| = 1$ denote the normalized LoS components, $\tilde{h}^{(s,d)},\tilde{h}_{k,n}^{(r,d)}\sim \mathcal{CN}(0,1)$ denote the normalized NLoS components, and $\kappa^{(s,d)} ,\kappa_{k}^{(r,d)} \ge 0$ denote the Rician factors. If $\kappa^{(s,d)}=0$ or $\kappa_{k}^{(r,d)} = 0$, the corresponding Rician fading reduces down to Rayleigh fading; if $\kappa^{(s,d)}\to \infty$ or $\kappa_{k}^{(r,d)} \to \infty$, only the fixed LoS component exists. For ease of analysis, we assume that any two IRSs are either aligned or sufficiently far from each other so that the channel between them is ignored as illustrated in Fig. 1. In contrast with \cite{3}, we assume that the instantaneous NLoS components (i.e., $\tilde{h}^{(s,d)}$ and $\tilde{h}_{k,n}^{(r,d)}$) vary slowly. Considering channel estimation overhead, assume that besides the fixed channel parameters (i.e., $\kappa^{(s,d)} ,\kappa_{k}^{(r,d)}, \bar{h}_{k,n}^{(s,r)}, \bar{h}^{(s,d)}, \bar{h}_{k,n}^{(r,d)}, \alpha_{k,n}^{(s,r)}, \alpha^{(s,d)},\alpha_{k}^{(r,d)}$), only the distributions of $\tilde{h}^{(s,d)},\tilde{h}_{k,n}^{(r,d)}$ are known at the source. 
	
	 Let $\bm{\theta}\triangleq(\theta_{k,n})_{k\in\mathcal{K},n\in\mathcal{N}_k}$ represent the phase shifts of the $K$ IRSs with $\theta_{k,n}$ being the phase shift of the $n$-th IRS element of the $k$-th IRS, where 
	\begin{equation}
	0\leq\theta_{k,n}< 2\pi,\ k\in\mathcal{K},n\in\mathcal{N}_k.
	\end{equation} 
	Considering practical implementation overhead, assume that the phase shifts $\bm{\theta}$ do not change with the instantaneous NLoS components.
	
	The receive signal at the destination can be expressed as 
	\begin{equation}
	\nonumber
	y(\bm{\theta})=\sqrt{P}\left(\underbrace{h^{(s,d)}+\sum_{k\in\mathcal{K},n\in\mathcal{N}_k}  h^{(r,d)}_{k,n}e^{j\theta_{k,n}}h_{k,n}^{(s,r)}}_{\triangleq h}\right)x+z,
	\end{equation}
	where $P$ denotes the transmit power, $x \sim \mathcal{CN}(0,1)$ is the transmitted signal, $h$ represents the equivalent channel between the source and the destination, $z\sim \mathcal{CN}(0,\sigma^2)$ represents the noise at the destination.	
	The channel capacity (in bit/s/Hz) of the multi-IRS-assisted system is given by 
	\begin{equation}
	C(\bm{\theta})=\log_2\left(\!\!1\!+\!\frac{P\left|h\right|^2}{\sigma^2}\right).
	\end{equation}
	Note the $C(\bm{\theta})$ is random due to the randomness of $\tilde{h}^{(s,d)}$, $\tilde{h}_{k,n}^{(r,d)}$, $k\in\mathcal{K}$, $n\in\mathcal{N}_k$. In the scenario where NLoS components vary slowly and are not known at the transmitter, the outage probability  is considered as the performance metric. For a given transmission  rate $R$ (in bit/s/Hz), the outage probability, defined as the probability that the outage event $C(\bm{\theta})< R$ happens, can be expressed as
	\begin{equation}
	\setlength{\abovedisplayshortskip}{2pt}
	P_{o}(\bm{\theta})\triangleq\Pr[C(\bm{\theta})< R]\\
	=\Pr\left[\left|h\right|^2\!\!< \frac{2^R-1}{SNR}\right],
	\setlength{\belowdisplayshortskip}{2pt}
	\end{equation}
	where $SNR\triangleq\frac{P}{\sigma^2}$ denotes the transmit SNR. 

	\section{analysis of outage probability}
	In this section, we analyze the outage probability $P_o(\bm{\theta})$. Define 
	\begin{equation}
	f\left(a,b,c\right)
	\triangleq e^{-\frac{a}{b}}\;\sum _{i=0}^{\infty }{\frac {(\frac{a}{b})^{i}}{i!}}\frac {\gamma (1+i,\frac{c}{b})}{\Gamma (1+i)},
	\end{equation}
	where $\gamma (p,q) \triangleq \int_0^q t^{p-1}e^{-t}dt$ is the lower incomplete gamma function, and $\Gamma (\mu) \triangleq (\mu-1)!$ is the gamma function.
	\begin{theorem}[Outage Probability]
	$P_{o}(\bm{\theta})$ is given by 
	\begin{equation}
	P_{o}(\bm{\theta})
	=f\left(g_{LoS}(\bm{\theta}),g_{NLoS},\frac{2^R-1}{SNR}\right), 
	\end{equation}

 	where $g_{LoS}(\bm{\theta}) $ is given by 
	\begin{equation}
 	g_{LoS}(\bm{\theta}) \triangleq
 	\bigg|\sqrt{\frac{\alpha^{(s,d)}\kappa^{(s,d)}}{\kappa^{(s,d)}+1}}\bar{h}_{s,d}+\sum_{k\in\mathcal{K},n\in\mathcal{N}_k}\sqrt{\frac{\alpha_{k}^{(r,d)}\alpha_{k}^{(s,r)}\kappa_{k}^{(r,d)}}{\kappa_{k}^{(r,d)}+1}}\bar{h}_{k,n}^{(r,d)}e^{j\theta_{k,n}} \bar{h}_{k,n}^{(s,r)}\bigg|^2,
 	\end{equation}
 	 and
	\begin{equation}
 	g_{NLoS}\triangleq\frac{\alpha^{(s,d)}}{\kappa^{(s,d)}+1}+\sum_{k\in\mathcal{K}}\frac{N_k\alpha_{k}^{(r,d)}\alpha_{k}^{(s,r)}}{\kappa_{k}^{(r,d)}+1}.
 	\end{equation}

	\end{theorem}
\begin{IEEEproof}
	By (3), we have $P_{o}(\bm{\theta}) = \Pr\left[|h|^2< \frac{2^R-1}{SNR}\right]$. Obviously, $h^{(s,d)} \sim \mathcal{CN}(\sqrt{\frac{\alpha^{(s,d)}\kappa^{(s,d)}}{\kappa^{(s,d)}+1}}\bar{h}_{s,d},\frac{\alpha^{(s,d)}}{\kappa^{(s,d)}+1})$, and $\sum_{k\in\mathcal{K},n\in\mathcal{N}_k}h^{(r,d)}_{k,n}e^{j\theta_{k,n}}h_{k,n}^{(s,r)}\sim \mathcal{CN}\Big(\sum_{k\in\mathcal{K},n\in\mathcal{N}_k}\sqrt{\frac{\alpha_{k}^{(r,d)}\kappa_{k}^{(r,d)}}{\kappa_{k}^{(r,d)}+1}}\bar{h}_{r,d,k}e^{j\theta_{k,n}} \bar{h}_{k,n}^{(s,r)},$ $\sum_{k\in\mathcal{K},n\in\mathcal{N}_k}\frac{\alpha_{k}^{(r,d)}\alpha_{k}^{(s,r)}}{\kappa_{k}^{(r,d)}+1}\Big)$. Thus $h \sim \mathcal{CN}\Big(\sqrt{\frac{\alpha^{(s,d)}\kappa^{(s,d)}}{\kappa^{(s,d)}+1}}\bar{h}_{s,d}+\sum_{k\in\mathcal{K},n\in\mathcal{N}_k}\sqrt{\frac{\alpha_{k}^{(r,d)}\alpha_{k}^{(s,r)}\kappa_{k}^{(r,d)}}{\kappa_{k}^{(r,d)}+1}}\bar{h}_{k,n}^{(r,d)}e^{j\theta_{k,n}} \bar{h}_{k,n}^{(s,r)},g_{NLoS}
	\Big)$. Thus,  $\frac{|h|^2}{g_{NLoS}}$ follows non-central chi-squared distribution with 2 degrees of freedom. Therefore, we complete the proof.
\end{IEEEproof}

	Note that $g_{LoS}(\bm{\theta})$ in (6) indicates the power of the LoS component of the equivalent channel $h$ and $g_{NLoS}$ in (7) indicates the power of the NLoS component of the equivalent channel $h$. For all $k\in \mathcal{K}$, $P_o(\bm{\theta})$ depends on $\alpha_{k}^{(r,d)}$ and $\alpha_{k}^{(s,r)}$ through their product $\alpha_{k}^{(r,d)}\alpha_{k}^{(s,r)}$, which represents the overall path loss of the reflected link via the $k$-th IRS. Furthermore, when $\alpha^{(s,d)}$, $\alpha_{k}^{(s,r)}\alpha_{k}^{(r,d)}$, $k\in\mathcal{K}$ increase by the same factor, $P_o(\bm{\theta})$ decreases.  

	Fig. 2 (a) and (b) plot the outage probability versus the phase shifts $\bm{\theta}$ and the transmit SNR, respectively. We see from Fig. 2 that the analytical curves (i.e., $P_o(\bm{\theta})$ in Theorem 1) are in excellent agreement with the numerical curves (i.e., Monte Carlo simulation results). Thus, Fig. 2 (a) and (b) verify  Theorem 1.  From Fig. 2, we can also see that $P_o(\bm{\theta})$ decreases with the SNR, and changes significantly with $\bm{\theta}$, which motivates us to minimize  $P_o(\bm{\theta})$ with respect to $\bm{\theta}$.
\begin{figure}[t]
	\centering
	\subfigure[$P_o(\bm{\theta})$ versus $\bm{\theta}$ at SNR = 15dB.]{
		\label{fig:subfig:a} 
		\includegraphics[width=3in]{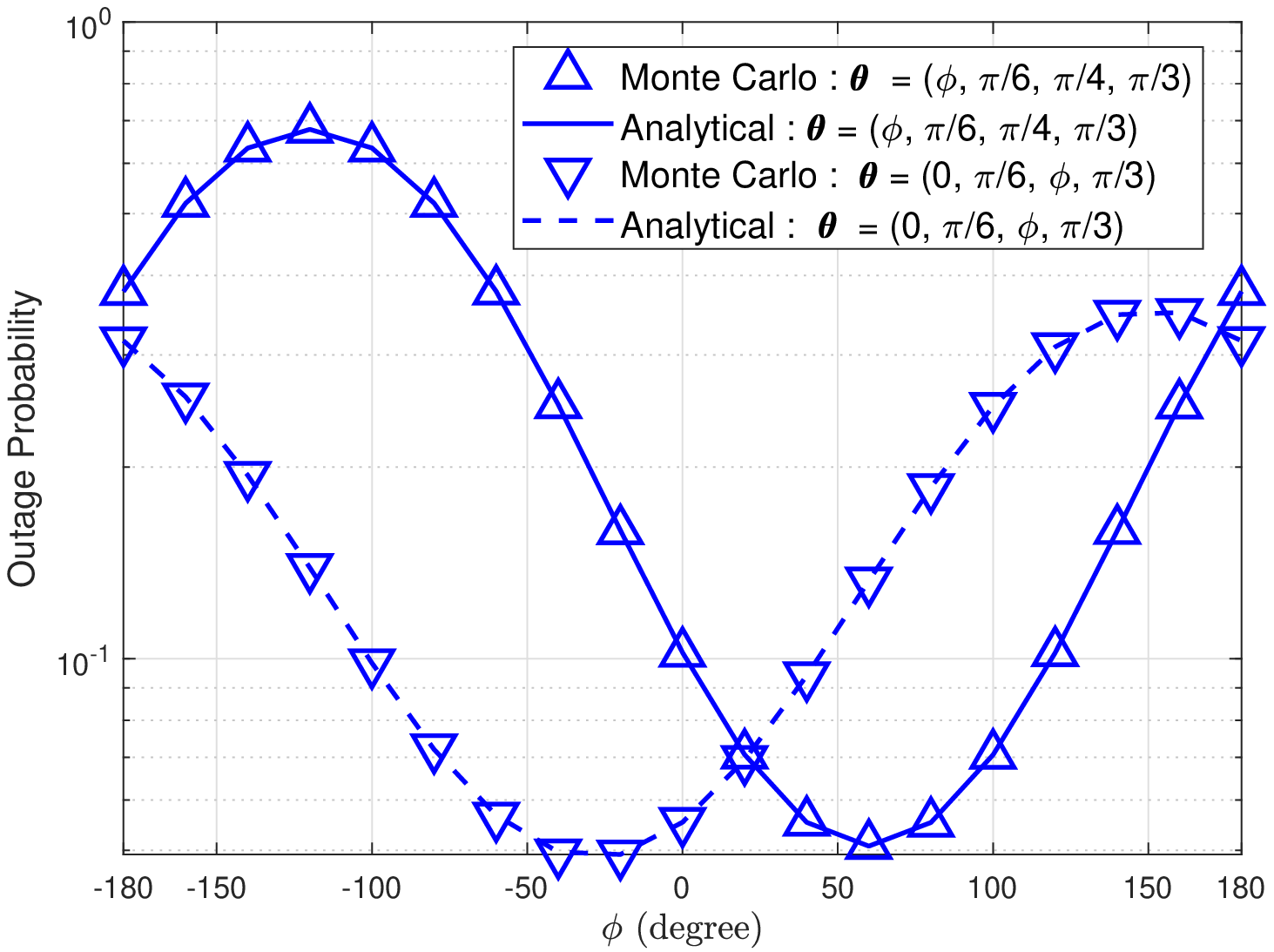}}
	\subfigure[$P_o(\bm{\theta})$ versus SNR.]{
		\label{fig:subfig:b} 
		\includegraphics[width=3in]{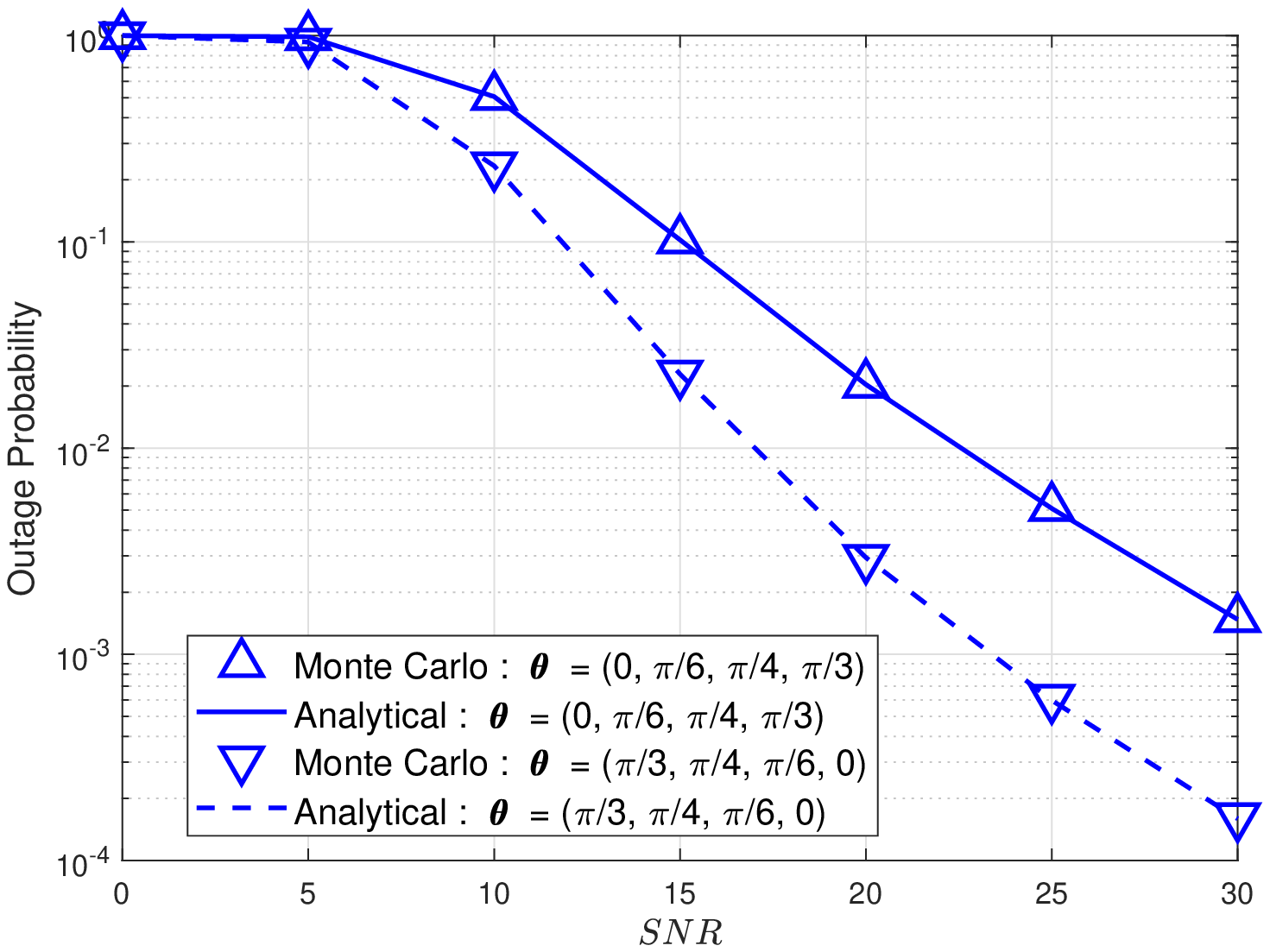}}
	\caption{Outage probability versus $\bm{\theta}$ and SNR. $K = 2$, $N_k = 2$, $k\in\mathcal{K}$, $R = 4$, $\alpha^{(s,d)} = 0.8$, $\alpha_{1}^{(s,r)} = \alpha_{2}^{(s,r)} = 1$, $\alpha_{1}^{(r,d)} = 0.6$, $\alpha_{2}^{(r,d)} = 0.1$, $\kappa^{(s,d)} = 2$, $\kappa_{1}^{(r,d)} = 10$, $\kappa_{2}^{(r,d)} = 15$.}
	\label{fig:subfig} 
\end{figure}

\section{optimization of outage probability}
 In this section, we minimize the outage probability $P_o(\bm{\theta})$ with respect to the phase shifts $\bm{\theta}$. First, the problem can be formulated as follows.
\begin{problem}[Outage Probability Optimization]
\begin{align}
\nonumber
P_o^* \triangleq &\min_{\bm{\theta}}\quad P_o(\bm{\theta}),\\
\nonumber
& \begin{array}{r@{\quad}r@{}l@{\quad}l}
s.t.& (1), 
\end{array}
\end{align}
where $P_o(\bm{\theta})$ is given by (5). Let $\bm{\theta}^*\triangleq (\theta_{k,n}^*)_{k\in \mathcal{K},n\in\mathcal{N}_k}$ denote an optimal solution of Problem 1.
\end{problem}

 Next, we solve Problem 1.

\begin{theorem}[Optimal Phase Shifts and Outage Probability]
	\begin{enumerate}[(i)]
		\item When $\kappa^{(s,d)}\ne 0$, any $\bm{\theta}$ satisfying (1) and  $\theta_{k,n}=\arg\frac{\bar{h}^{(s,d)}}{\bar{h}^{(r,d)}_{k,n}\bar{h}_{k,n}^{(s,r)}} $ for all $k \in \mathcal{K}$ with $\kappa_{k}^{(r,d)} \ne 0$ is optimal.
		\item When $\kappa^{(s,d)} = 0$, any  $\bm{\theta}$ satisfying (1) and  $\theta_{k_1,n_1}-\theta_{k_2,n_2}=\arg\frac{\bar{h}^{(r,d)}_{k_1,n_1}\bar{h}_{k_1,n_1}^{(s,r)}}{\bar{h}^{(r,d)}_{k_2,n_2}\bar{h}_{k_2,n_2}^{(s,r)}} $ for all $k_1,k_2 \in\mathcal{K}$ with $\kappa_{k_1}^{(r,d)} \ne 0$ and $\kappa_{k_2}^{(r,d)} \ne 0$ is optimal.		
		\item $P_o^* = f\left(g_{LoS}^*, g_{NLoS}, \frac{2^R-1}{SNR}\right)$, where 
	\end{enumerate}
	\begin{equation}
	\begin{aligned}
	g_{LoS}^*\triangleq \left(\!\!\!\sqrt{\frac{\alpha^{(s,d)}\kappa^{(s,d)}}{\kappa^{(s,d)}+1}}\!+\!\!\!\sum_{k\in\mathcal{K}}\!\!N_k\sqrt{\frac{\alpha_{k}^{(r,d)}\alpha^{(s,r)}_k\kappa_{k}^{(r,d)}}{\kappa_{k}^{(r,d)}+1}}\right)^2\!.
	\end{aligned}
	\end{equation}

\end{theorem}
\begin{IEEEproof}
	By (4), we have $f_{a}'(a,b,c)=
	\frac{1}{b}e^{-\frac{a}{b}}\left(\sum_{i=0}^{\infty}\frac{(\frac{a}{b})^i}{i!}\left(Q(4+2i,\frac{2c}{b}) - Q(2+2i,\frac{2c}{b})\right)\right)$, where $Q(m, t)$ denotes the C.D.F. of the central chi-squared distribution with $m$ degrees of freedom, evaluated at $t$. As $Q(m, t)$ is strictly decreasing in $m$ for all $t$, we have $Q(4+2i,\frac{2c}{b}) < Q(2+2i,\frac{2c}{b})$. Thus, $f_{a}'(a,b,c)<0$, i.e., $f(a,b,c)$ is monotonically decreasing in $a$. 
	Therefore, Problem 1 is equivalent to 	
	\begin{align*}
	g_{LoS}^*\triangleq&\max_{\bm{\theta}}\quad g_{LoS}(\bm{\theta}),\\
		& \begin{array}{r@{\quad}r@{}l@{\quad}l}
	s.t.& (1).
	\end{array}
	\end{align*}
	When $\kappa^{(s,d)} \ne 0$, for all $k \in \mathcal{K}$ with $\kappa_{k}^{(r,d)} \ne 0$, by triangle inequality, $g_{LoS}(\bm{\theta}) \le  g_{LoS}^*$, where the equality holds when $\arg \bar{h}^{(s,d)} = \arg \bar{h}^{(r,d)}_{k,n}e^{j\theta^*_{k,n}} \bar{h}_{k,n}^{(s,r)}$. Thus, we can show statement (i).
	When $\kappa^{(s,d)} = 0$, for all  $k \in \mathcal{K}$ with  $\kappa_{k}^{(r,d)} \ne 0$, by triangle inequality, $g(\bm{\theta}) \le  \bigg(\sum_{k\in\mathcal{K}}N_k\sqrt{\frac{\alpha_{k}^{(r,d)}\alpha_{k}^{(s,r)}\kappa_{k}^{(r,d)}}{\kappa_{k}^{(r,d)}+1}}\bigg)^2$, where the equality holds when $\arg{\bar{h}^{(r,d)}_{k_1,n_1}e^{j\theta_{k_1,n_1}^*}\bar{h}_{k_1,n_1}^{(s,r)}}=\arg{\bar{h}_{k_2,n_2}^{(r,d)}e^{j\theta_{k_2,n_2}^*}\bar{h}_{k_2,n_2}^{(s,r)}}$ for all $k_1\ne k_2, n_1 \ne n_2, k_1,k_2 \in\mathcal{K},n_1,n_2\in\mathcal{N}_k$. Thus, we can show the statement (ii).
	From the above analysis, we can obtain $g_{LoS}^*$ and $P_o^*$. 
\end{IEEEproof}

Note that $g_{LoS}^*$ in (8) indicates the maximum achievable power of the LoS component of the equivalent channel $h$. Theorem 2 (i) indicates that when the LoS component of the direct link exists, the phase changes over all reflected links with LoS components should be aligned with that of the direct link. Theorem 2 (ii) indicates that when the direct link does not have an LoS component, the phase changes over all reflected links with an LoS component should be aligned. 

Next, we characterize the impacts of $K$ and $N_k$, $k\in\mathcal{K}$ on $P_o^*$.
\begin{lemma}[Properties]
	If $\kappa_{k}^{(r,d)}>1$ for all $k\in\mathcal{K}$, $P_o^*$ decreases with $K$ and with $N_k$, $k\in \mathcal{K}$.
\end{lemma}
\begin{IEEEproof}
 By (7) and (8), we can easily show that  when $\kappa_k^{(r,d)}>1$, 
 $k\in\mathcal{K}$, $g_{NLoS}$, $g^*_{LoS}$ and $\frac{g^*_{LoS}}{g_{NLoS}}$ increase with $K$ and with $N_k$, $k\in \mathcal{K}$. In addition, the C.D.F. of a non-central chi-squared distribution decreases with the noncentrality parameter and increases with the upper limit of the value of the random variable.  Thus, we can prove Lemma 1.
\end{IEEEproof}

Lemma 1 indicates that $P_o^*$ decreases with the number of IRSs and with the number of elements of each IRS, if the power of each LoS component of the channel between each IRS and the destination is larger than the average power of the corresponding NLoS component. This can be seen from Fig. 3 (a) and (b). In addition, from Fig. 3 (a) and (b), we see that the optimal outage probability of the multi-IRS-assisted system is much smaller than that of the system without IRSs, especially at large $K$ and $N_k$, $k\in\mathcal{K}$. 
\begin{figure}[t]
	\centering
	\subfigure[$P_o^*$ versus $K$ at $N_k = 20$, $ \alpha_k^{(s,r)} = \alpha_k^{(r,d)} = 0.01$, $\kappa_k^{(r,d)} = 10$, $k\in\mathcal{K}$.]{
		\label{fig:subfig:a} 
		\includegraphics[width=3in]{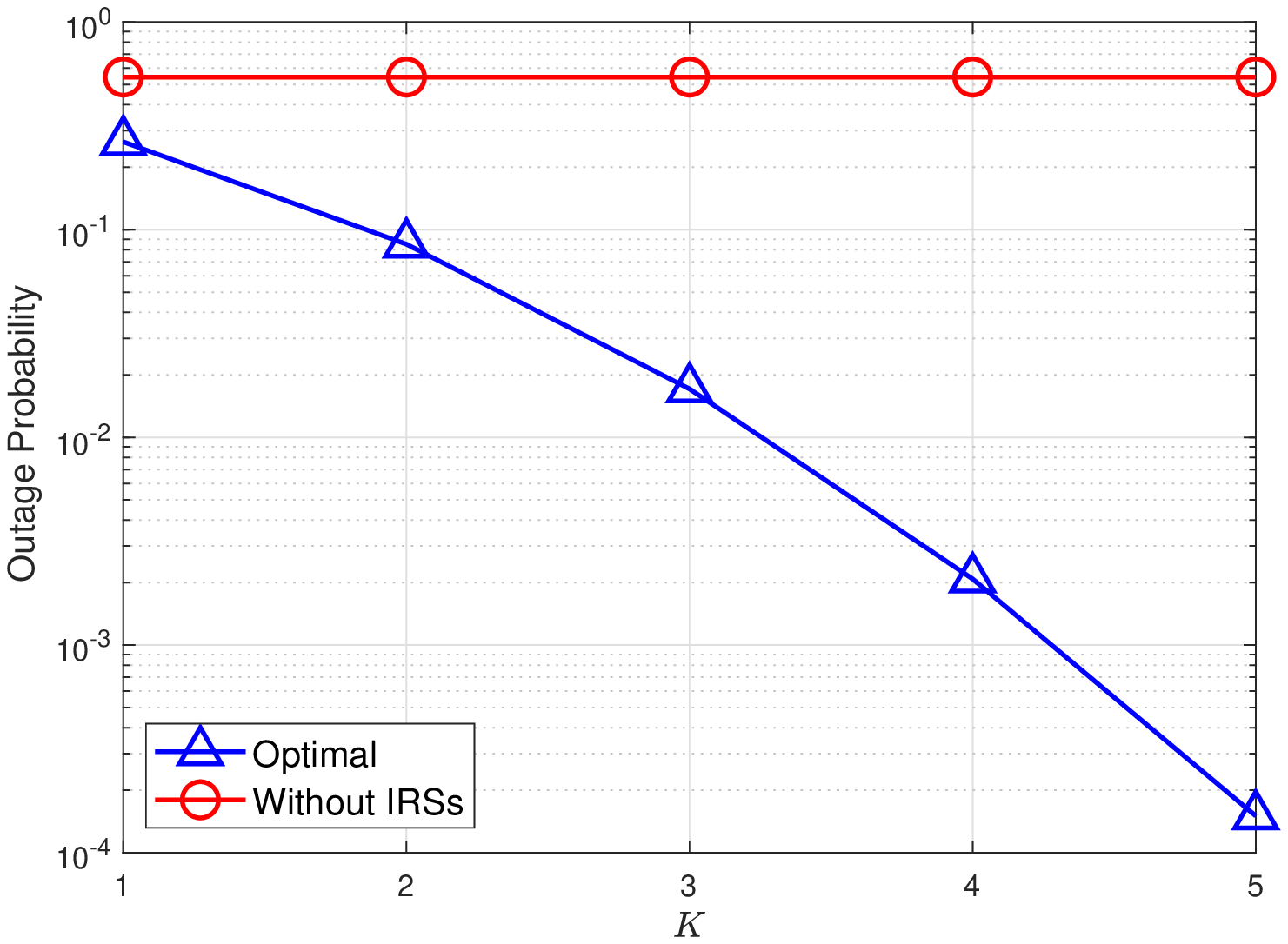}}
	\subfigure[$P_o^*$ versus $N_k$, $k\in\mathcal{K}$ at $K = 2$, $\alpha_{1}^{(s,r)}=\alpha_{1}^{(r,d)} = 0.01$, $\alpha_{2}^{(s,r)} =\alpha_{2}^{(r,d)} = 0.05$, $\kappa_1^{(r,d)} = 10$, $\kappa_2^{(r,d)} = 15$.]{
		\label{fig:subfig:b} 
		\includegraphics[width=3in]{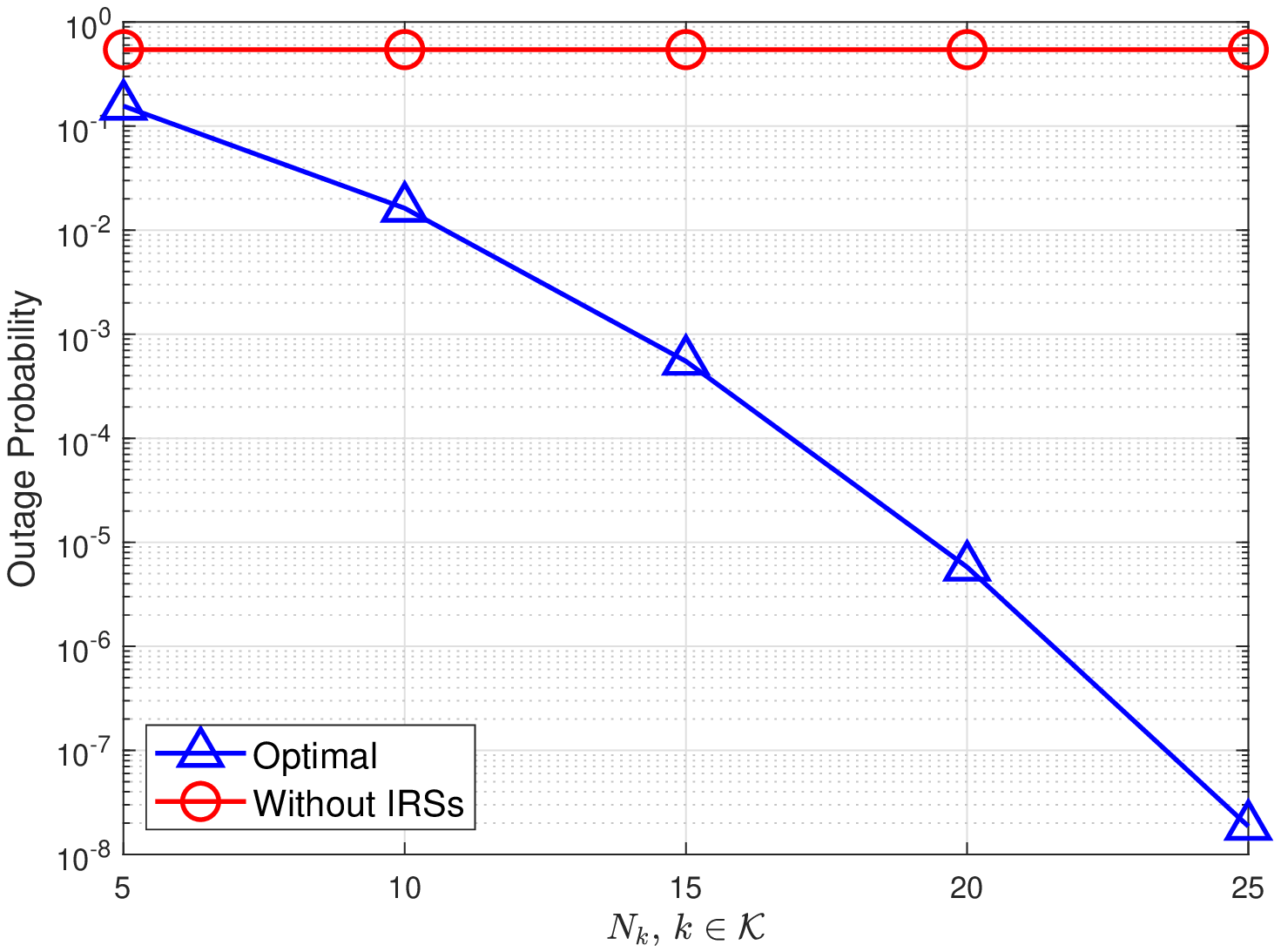}}
	\caption{Optimal outage probability versus  $K$ and $N_k$, $k\in\mathcal{K}$. $R = 4$, $\alpha^{(s,d)} = 0.5$, $\kappa^{(s,d)} = 3$, $SNR = 15$dB.}
	\label{fig:subfig} 
\end{figure}

\section{Asymptotically optimal outage}
In this section, we characterize the asymptotic behavior of the optimal outage probability $P_o^*$ at high SNR. For ease of exposition, in this section, we explicitly write the optimal outage probability as a function of SNR, i.e., $P_o^*(SNR)$.

\begin{lemma}[Asymptotically Optimal Outage Probability]
	$P^{*}_o(SNR)\overset{SNR \to \infty}{\sim}\tilde{P}_{o}^*(SNR)$, 
   	where $\tilde{P}_{o}^*(SNR)$ is given by $\tilde{P}_{o}^*(SNR)=\frac{2^R-1}{g_{NLoS}SNR}
   	\exp\left(-\frac{g^*_{LoS}}{g_{NLoS}}\right)$.\footnote[3]{$f(\xi)\overset{\xi \to \infty}{\sim} g(\xi)$ means $\lim_{\xi \to \infty}\frac{f(\xi)}{g(\xi)} = 1$.}

\end{lemma}

\begin{IEEEproof}
	As
	\begin{equation}
	\nonumber
	\lim_{SNR \to \infty} \frac{\gamma(1+i,\frac{2^R-1}{bSNR})}{\frac{(\frac{2^R-1}{bSNR})^{i+1}}{i+1}} \overset{(a)}{=} \lim_{SNR \to \infty} e^{-\frac{2^R-1}{bSNR}}= 1,
	\end{equation}
	where $(a)$ is due to L’Hopital’s Rule, we have 	
	\begin{align*}
		\nonumber
	\lim_{SNR \to \infty}\frac{f(a,b,\frac{2^R-1}{SNR})}{\frac{2^R-1}{SNR}e^{-\frac{x}{y}}}&=\sum _{i=0}^{\infty }\lim_{SNR \to \infty}\frac{e^{-\frac{a}{b}}{\frac {(\frac{a}{b})^{i}}{i!}}\frac {\gamma (1+i,\frac{c}{b})}{\Gamma (1+i)}}{e^{-\frac{a}{b}}\frac{2^R-1}{SNR}}= 1.
	\end{align*}
	Thus, by Theorem 2 (iii), we complete the proof.
\end{IEEEproof}	

Lemma 2 indicates that $P_o^*(SNR)$ is inversely proportional to the SNR in the high SNR regime. Fig. 4 (a) shows the asymptotically optimal outage probability versus the transmit SNR. We can see from Fig. 4 (a) that when SNR increases, the gap between each analytical curve (plotted using $P_o^*$ given in Theorem 2 (iii)) and the corresponding asymptotic curve (plotted using $\tilde{P}_o^*(SNR)$ given in Lemma 2) decreases, verifying Lemma 2. Furthermore, we also see that $P_o^*(SNR)$ in the high SNR regime and $\tilde{P}_o^*(SNR)$ are  inversely proportional to the SNR.

Next, we characterize the impacts of $ \kappa^{(s,d)}, \kappa_{k}^{(r,d)},k\in \mathcal{K}$ on $\tilde{P}_{o}^*(SNR)$ in the high SNR regime.
\begin{lemma}[Asymptotic Properties]
	$\tilde{P}_{o}^*(SNR)$ decreases with $ \kappa^{(s,d)}, \kappa_{k}^{(r,d)},k\in \mathcal{K}$.
\end{lemma}

\begin{IEEEproof}
	We can easily show that $\frac{\partial \tilde{P}_{o}^*(SNR)}{\partial \kappa^{(s,d)}}<0$ and $\frac{\partial \tilde{P}_{o}^*(SNR)}{\partial \kappa_{k}^{(r,d)}}<0, k\in \mathcal{K}$. The details are omitted due to page limitation. 
\end{IEEEproof}

Lemma 3 indicates that $\tilde{P}_{o}^*(SNR)$ decreases with the power of each LoS component, as shown in Fig. 4 (b). In addition, from Fig. 4 (b), we can observe that $\tilde{P}^*_o(SNR)$ is much smaller than the asymptotic outage probability of the system without IRSs, especially at large $\kappa_k^{(r,d)}, k\in\mathcal{K}.$ 

\begin{figure}[t]
	\centering
	\subfigure[$\tilde{P}_{o}^*$ versus SNR at $\kappa_k^{(r,d)} = 10$, $k\in\mathcal{K}$.]{
		\label{fig:subfig:a} 
		\includegraphics[width=3in]{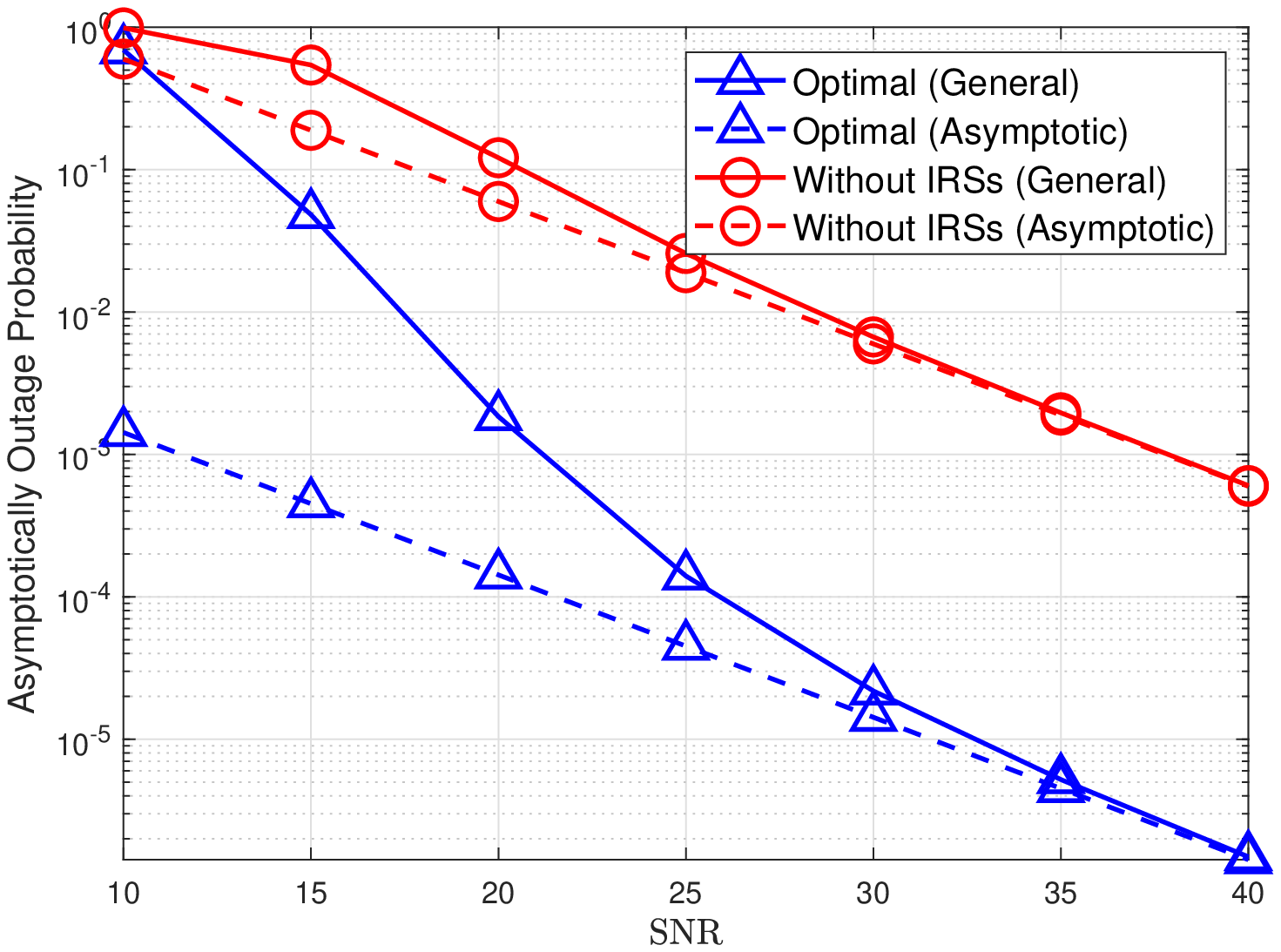}}
	\subfigure[$\tilde{P}_{o}^*$ versus $\kappa_k^{(r,d)}$ at SNR = 40dB. ]{
		\label{fig:subfig:b} 
		\includegraphics[width=3in]{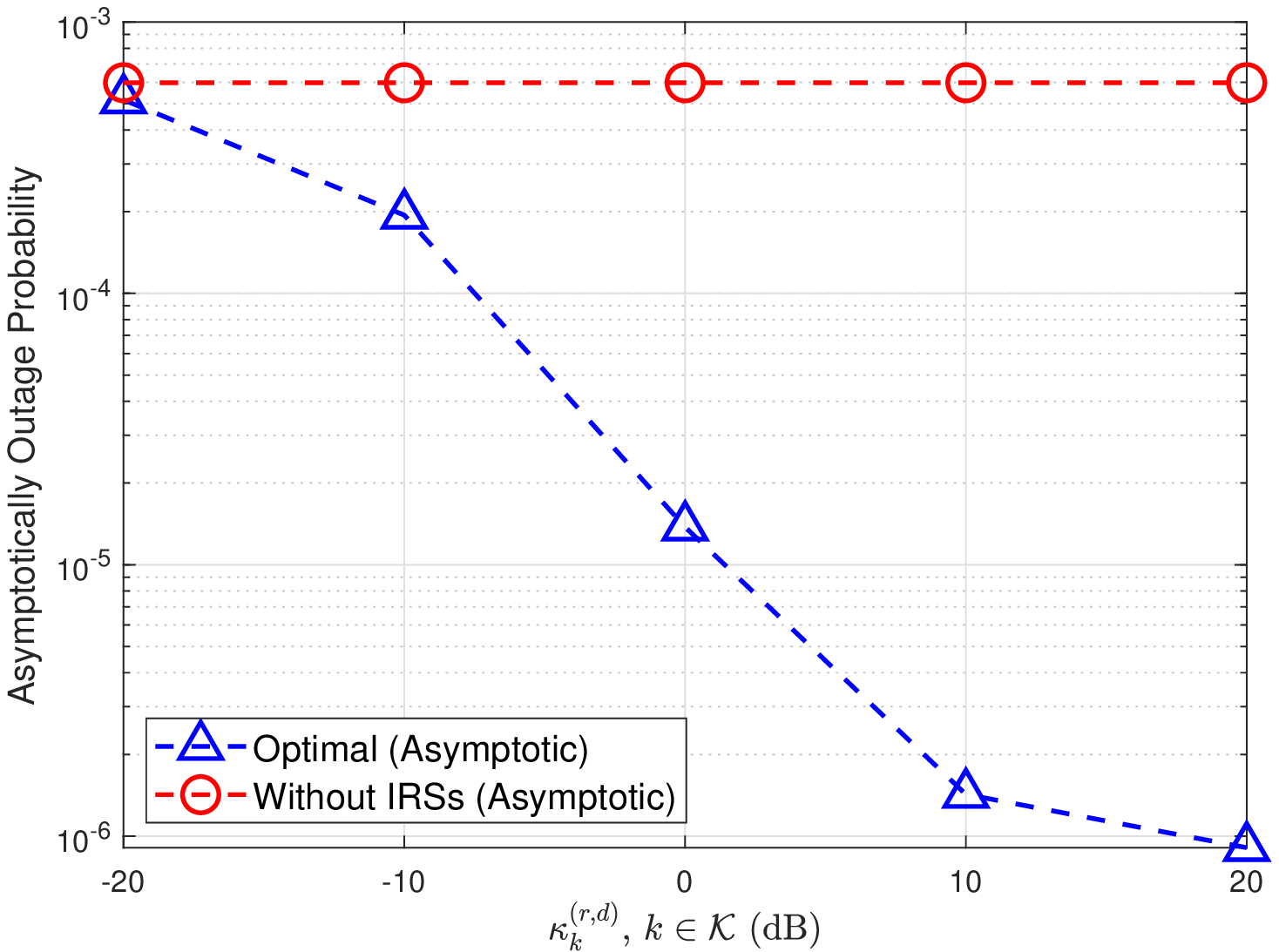}}
	\caption{Asymptotically optimal outage probability versus $SNR$ and $\kappa_k^{(r,d)}$, $k\in\mathcal{K}$. $K = 2$, $N_k = 8$, $R = 4$, $\alpha^{(s,d)} = 0.5$, $\alpha_{1}^{(s,r)}=\alpha_{1}^{(r,d)} = 0.01$, $\alpha_{2}^{(s,r)} =\alpha_{2}^{(r,d)} = 0.05$, $\kappa^{(s,d)} = 3$.}
	\label{fig:subfig} 
\end{figure}
\section{conclusion}
In this letter, we investigated a multi-IRS-assisted system under Rician fading where the phase shifts adapt to only the LoS components. First, we obtained the expression of the outage probability which is a function of the phase shifts of all IRSs. Then, we obtained the closed-form optimal phase shifts that minimize the outage probability. Next, we obtained the expression of the asymptotically optimal outage probability in the high SNR regime. Finally, we characterized the impacts of the number of IRSs, the number of elements of each IRS, and the power of each LoS component on the optimal outage probability.
\bibliographystyle{IEEEtran}

\end{document}